\documentclass[12pt,preprint]{aastex}
\usepackage{natbib}
\shorttitle{Surface Waves}
\shortauthors{M. Roth et al.}

\begin{document}


\title{Surface waves in solar granulation observed with {\sc Sunrise}}


\author{\textsc{
M.~Roth,$^{1}$
M.~Franz,$^{1}$
N.~Bello Gonz\'alez,$^{1}$
V.~Mart\'inez Pillet,$^{2}$
J.~A.~Bonet,$^{2}$
A.~Gandorfer,$^{3}$
P.~Barthol,$^{3}$
S.~K.~Solanki,$^{3,7}$
T.~Berkefeld,$^{1}$
W.~Schmidt,$^{1}$
J.~C.~del~Toro~Iniesta,$^{4}$
V.~Domingo,$^{5}$
M.~Kn\"olker$^{6}$
}}

\affil{
$^{1}$Kiepenheuer-Institut f\"ur Sonnenphysik, Sch\"oneckstr. 6, 79104 Freiburg, Germany.\\
$^{2}$Instituto de Astrof\'{\i}sica de Canarias, C/Via L\'actea s/n, 38200 La Laguna, Tenerife, Spain.\\
$^{3}$Max-Planck-Institut f\"ur Sonnensystemforschung, Max-Planck-Str. 2, 37191 Katlenburg-Lindau, Germany.\\
$^{4}$Instituto de Astrof\'{\i}sica de Andaluc\'{\i}a (CSIC), Apartado de Correos 3004, 18080 Granada, Spain.\\
$^{5}$Grupo de Astronom\'{\i}a y Ciencias del Espacio, Universidad de Valencia, 46980 Paterna, Valencia, Spain.\\
$^{6}$High Altitude Observatory, National Center for Atmospheric Research, P.O. Box 3000, Boulder CO 80307-3000, USA.\\
$^{7}$School of Space Research, Kyung Hee University, Yongin, Gyeonggi, 446-701, Korea.
}

\email{mroth@kis.uni-freiburg.de}

\begin{abstract}
Solar oscillations are expected to be excited by turbulent flows in the intergranular lanes near the solar surface.
Time series recorded by the IMaX instrument aboard the {\sc Sunrise} observatory reveal solar oscillations at high resolution, which allow studying the properties of oscillations with short wavelengths.  
We analyze two times series with synchronous recordings of Doppler velocity and continuum intensity images with durations of 32\thinspace min and 23\thinspace min, resp., recorded close to the disk center of the Sun to study the propagation and excitation of solar acoustic oscillations.
In the Doppler velocity data, both the standing acoustic waves and the short-lived, high-degree running waves are visible. The standing waves are visible as temporary enhancements of the amplitudes of the large-scale velocity field due to the stochastic superposition of the acoustic waves. We focus on the high-degree small-scale waves by suitable filtering in the Fourier domain.
Investigating the propagation and excitation of $f$- and $p_1$-modes with { wave numbers $k > 1.4$\thinspace 1/Mm} we find that also exploding granules contribute to the excitation of solar $p$-modes in addition to the contribution of intergranular lanes.
\end{abstract}



\section{Introduction}
\label{sec:intro}

Solar $p$-modes are believed to be stochastically excited in the near surface layers of the Sun by turbulent convection. 
The process can be described as acoustic radiation  by turbulent multipole sources~\citep{unno64}.
Further theoretical descriptions of the excitation of solar oscillations follow the work on stochastic driving by turbulent convection presented by~\cite{goldreich77} and~\cite{goldreich88}. These models were successful in explaining the acoustic power distribution over a wide range in frequencies as is manifested in the ensemble of solar $p$-modes.

Numerical simulations of convection, as described e.g. by~\cite{nordlund85} or~\cite{stein01}, predict supersonic downdrafts which could play a role in the excitation of the solar oscillations. Actually, the turbulent velocity field from the numerical simulations by~\cite{stein01} were used by~\cite{samadi03} to determine the energy supply rate of convection into the solar oscillations.

For the seismology of the Sun and sunlike stars, knowledge of the stochastic properties of the source function that drives $p$-modes would be a valuable input. In local helioseismology, e.g., the sensitivity functions which connect the acoustic properties of the $p$-modes with physical quantities in the Sun depend on the source function whose properties are not known and must be estimated~\citep{birch04}.
In addition, the Sun could serve as a paradigm for stellar seismology since the interaction of pulsations and convection and the energy gain and loss of $p$-modes from radiation and convection is a general problem of stellar physics~\citep{houdek06}. 

When searching for the possible locations of the wave sources on the Sun, the turbulent downdrafts in the dark intergranular lanes are primary targets. 
So-called ``acoustic events'', which are individual sunquakes with epicenters near the solar surface and located in the intergranular lanes, are assumed to feed continuously energy into the resonant $p$-modes of the Sun.
Studies of such acoustic events were carried out by~\cite{rimmele95} and~\cite{goode98} with the result that such events have typical durations of 5 minutes and are caused by collapsing intergranular lanes.

In this contribution our basis for further studies of the driving of the solar oscillations are Dopplergrams of the velocity above the solar surface. Time series recorded with the IMaX instrument~\citep{Pillet2010} aboard the balloon-borne observatory {\sc Sunrise}~\citep{Solanki2010, Barthol2010} provide the spatial and temporal resolution needed to measure the propagation of acoustic oscillations in the granules and the intergranular lanes. We assume that acoustic events could be observed as they eventually cause waves to propagate away from the location of the source. The unprecedented spatial resolution of the data makes it worthwhile to study the excitation and propagation of acoustic waves visible in the solar photosphere.
As a new result we find that the acoustic power is not only enhanced in the intergranular lanes and along the borders of the granules but also above large granules, where splitting processes are about to occur. 
Surprisingly, the excitation of waves of high wave number seem to be a diagnostic of pending granule splitting.


\section{Observation and data reduction}
\label{sec:obs}
For our study we use a time series of solar granulation with a field of view of 51$\arcsec$.5 by 51$\arcsec$.5 located close to disk center. The data were obtained with the Imaging Magnetograph eXperiment -- IMaX -- ~\citep{Pillet2010} onboard the {\sc SUNRISE} balloon-borne solar observatory~\citep{Solanki2010, Barthol2010} between
00:35:49~UT and 00:59:57~UT (1st series) and
01:30:41~UT and 02:02:48~UT on June 9, 2009 (2nd series).
The IMaX instrument and the reduction process which was applied to the data is described in detail in~\cite{Pillet2010} and shall be briefly summarized in the following.

The IMaX filtergraph was operated in the V5-6 observing mode, recording all Stokes parameters at $\lambda=[\pm 80,~\pm40,~+227]$~m\AA\ with respect to the line center of \ion{Fe}{1} 5250.2~\AA\ ($g_{\rm eff} = 3$, $\chi = 0.121$~eV). The exposure time for a picture at a single wavelength position was 6~s, while the entire line was scanned in 33~s.

In the data reduction process, the dark current was removed, and a flat field procedure damped residual impurities, i.e. dust particles or scratches. In addition, the interference fringes -- caused by the polarizing beam splitter -- were removed from the images by attenuating their contribution in Fourier space. Furthermore, a spatially dependent blue-shift over the FOV was calibrated and removed. Phase diversity (PD) measurements that were recorded close in time to the actual data, were used to reconstruct all pictures. The corrected images of the two cameras were merged to increase the S/N ratio which is of the order of {$4 \times 10^{2}$} {related to the continuum intensity ($I_c$)}in the reconstructed data. In a last step, crosstalk induced by residual motion was removed. Thus, we obtained time series with lengths of 24 and 32~minutes, resp., and a cadence of 33~s. A spatial resolution between $0.\arcsec15$ -  $0.\arcsec18$ was achieved by combining the on-board image stabilization system~\citep{gandorfer10,berkefeld10,Pillet2010}. 

Besides Doppler velocity ($v_{Dopp}$) and continuum intensity ($I_c$), which we will use for further investigation, other quantities like line center intensity ($I_{center}$), linear and circular polarization, and line width (FWHM) were computed. The quantities $v_{Dopp}$, $I_{center}$, and FWHM are inferred from a Gaussian fit to the four points in the line after they have been normalized using the continuum intensity at $+$227~m\AA. The velocity scale has been calibrated assuming a convective blue shift of 0.2~km\,s$^{-1}$ at disk center, and the absolute error in $v_{Dopp}$ was estimated to be of the order of $\pm 0.2 $~km\,s$^{-1}$.


\section{Solar Oscillations in IMaX Data}
\subsection{Acoustic Power}
\label{sec:power}
\begin{figure}[h]
\plotone{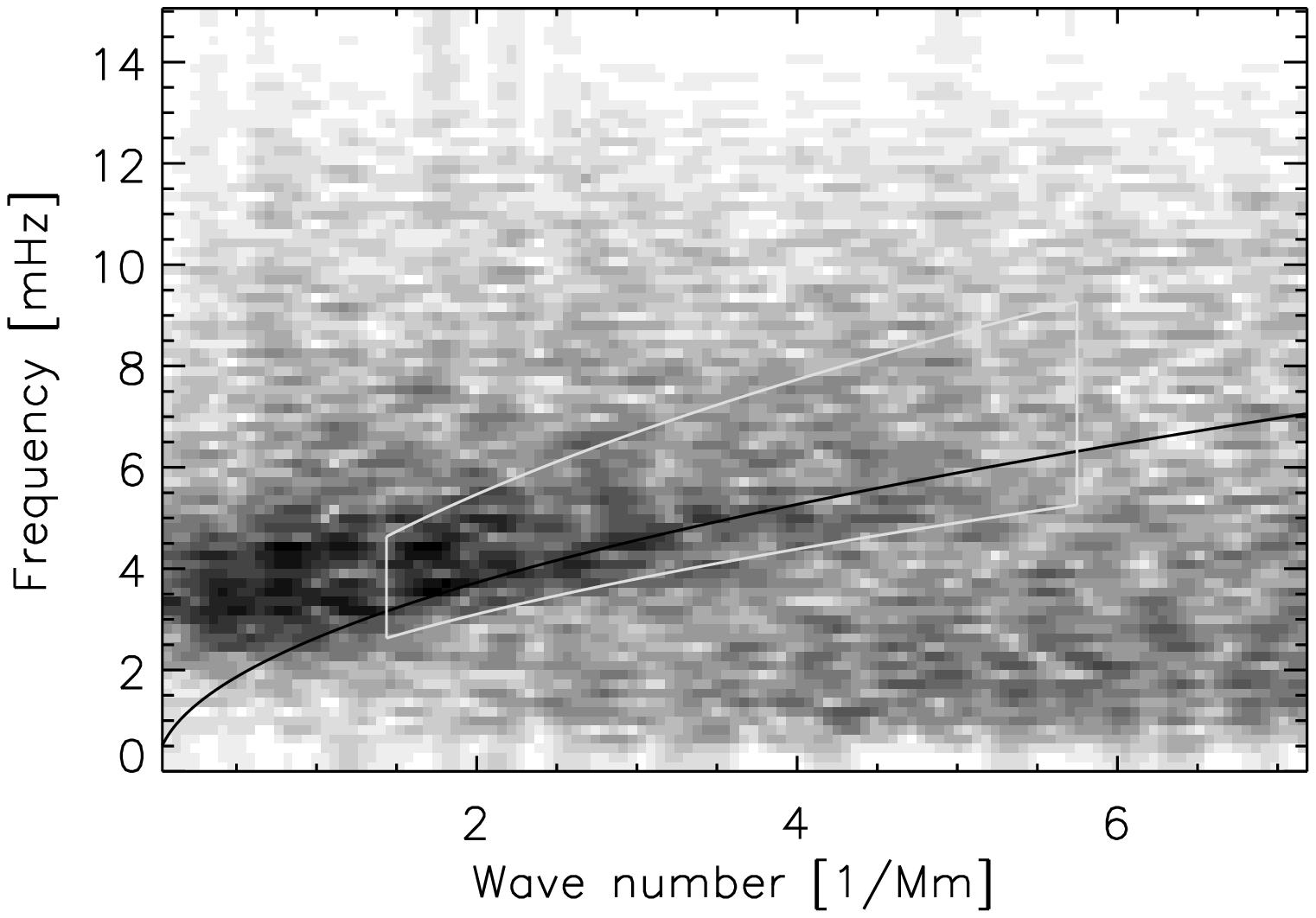}
\caption{Spectral power density of the measured Doppler velocity as a function of {wave number} and frequency in arbitrary units. The color code reflects the mode energy: large amplitudes are coded dark, amplitudes compatible with zero are coded white.
The black solid line indicates the theoretical position of the $f$-mode. The gray solid lines mark the borders of a Hanning band pass filter, which is applied to the data. 
\label{fig:power}}
\end{figure}

The two time series of the Doppler velocity measured by IMaX reveal a ``flying shadow'' pattern with a characteristic time scale of five minutes and a characteristic spatial length scale between 5$\arcsec$ and 10$\arcsec$. Such a movie will be part of the electronic supplemenatary material of the introductory letter by~\cite{Solanki2010}. This pattern is due to the solar $f$- and $p$-modes present on the Sun. The ``flying shadows'' originate from the stochastic superposition of a multitude of solar acoustic eigenmodes of various spatial wavelength, temporal frequency, and amplitude, which cause these ``Five-Minute Oscillations''.
For a quantitative analysis we investigate the power distribution in the diagnostic diagram, where the spectral power density is displayed as a function of {wave number $k$} and temporal frequency $\nu$, cf. Fig.~\ref{fig:power}.
The frequency resolution in the Fourier domain is 0.52\thinspace mHz.
We find a significant enhancement of oscillatory power between 2 and 7\thinspace mHz in the $f$- and $p$-mode area. Ridges of solar acoustic oscillations are at best visible only in outlines at high {wave numbers}. 
The $f$-mode ridge seems to extend until a {wave number of  $\approx 5$\thinspace Mm$^{-1}$}, where the ridge is cut by the acoustic cut-off frequency at $\approx$ 5.3\thinspace mHz (see Fig.~\ref{fig:power}). The velocity of the superposition of $f$-modes has an rms-amplitude of 147\thinspace m/s with a maximum velocity of 902\thinspace m/s.  
{From the location of enhanced power in this diagram we conclude that besides the superposition of large-scale standing acoustic waves with low wave numbers also the small-scale waves with high wave numbers are excited. These small-scale waves are expected to have a shorter lifetime than the large-scale waves.} In the following we will focus on the study of these small-scale oscillations.
Before doing so, we would like to note that besides the sonic power we find also subsonic power below the $f$-mode.
These subsonic velocities correspond to the convective granular motions we observe.


\subsection{Filter}
\label{sec:filter}
We study the excitation and propagation of solar $p$-modes by first filtering the data.
In order to remove trends in the time series a ``first-difference'' filter is applied. Further filtering for the $f$- and $p_1$-modes is carried out in the Fourier domain.
A fast Fourier transformation is applied to the time series. The result of this is multiplied by a filter that acts as a band-pass including the $f$-mode and the $p_1$-mode ridges. Figure~\ref{fig:power} outlines (gray lines) this three-dimensional filter as projection onto the $l-\nu$ diagram. 
The filter is based on Hanning windows with increasing widths as a function of the absolute magnitude of the wavenumber $k$.
Further filter properties are that the filter allows only those modes with {wave numbers $1.4$\thinspace Mm$^{-1}\le l \le 5.7$\thinspace Mm$^{-1}$} to pass, i.e. the filter removes the contributions of large-scale waves and subsonic granular motions.
After filtering, the data is transformed back into the time domain. 
The resulting time series contains only the small-scale oscillatory motions.


\subsection{Excitation and Propagation of Running Waves}
\label{sec:running}
\begin{figure*}[h]
\epsscale{0.8}
\plotone{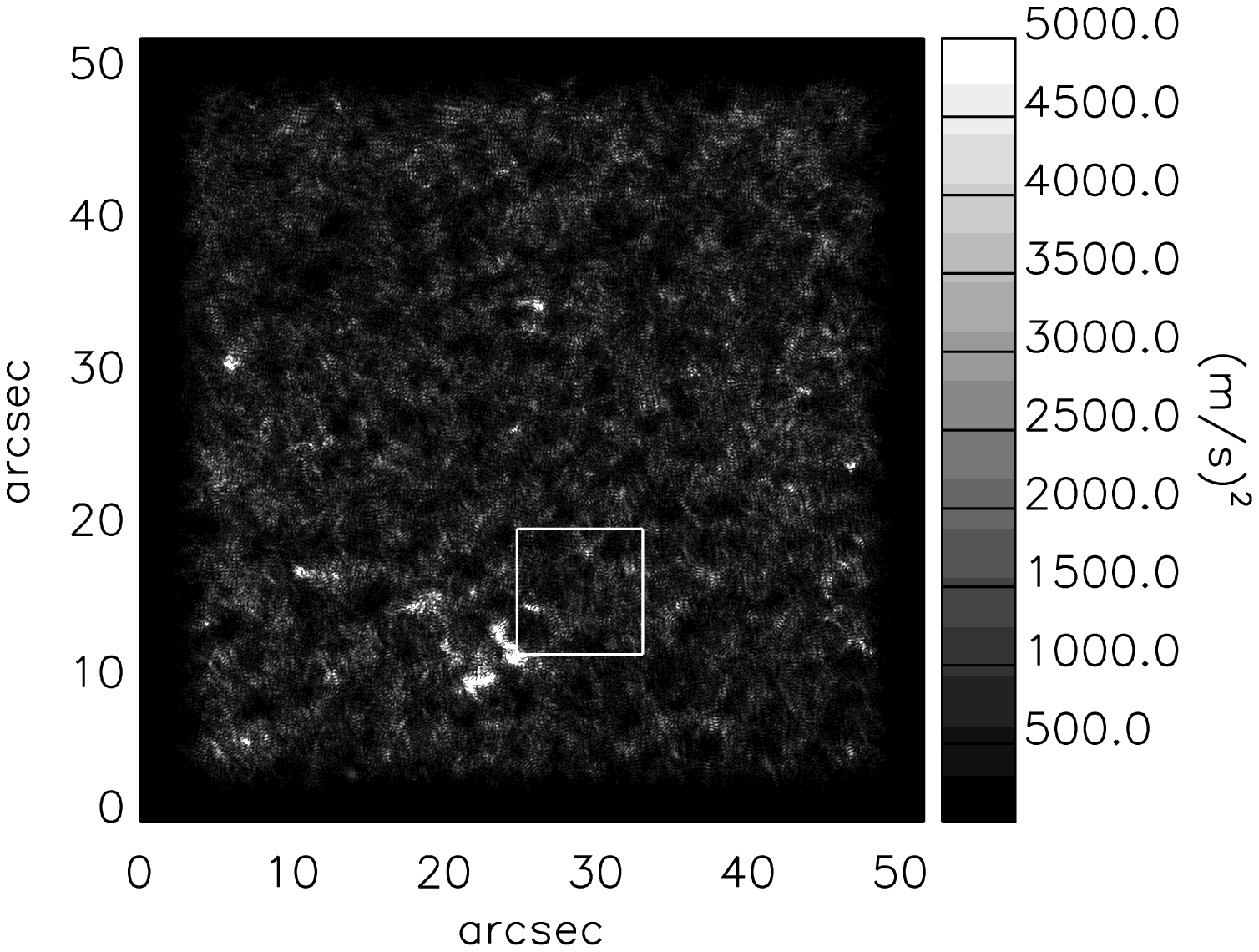}\\
\plotone{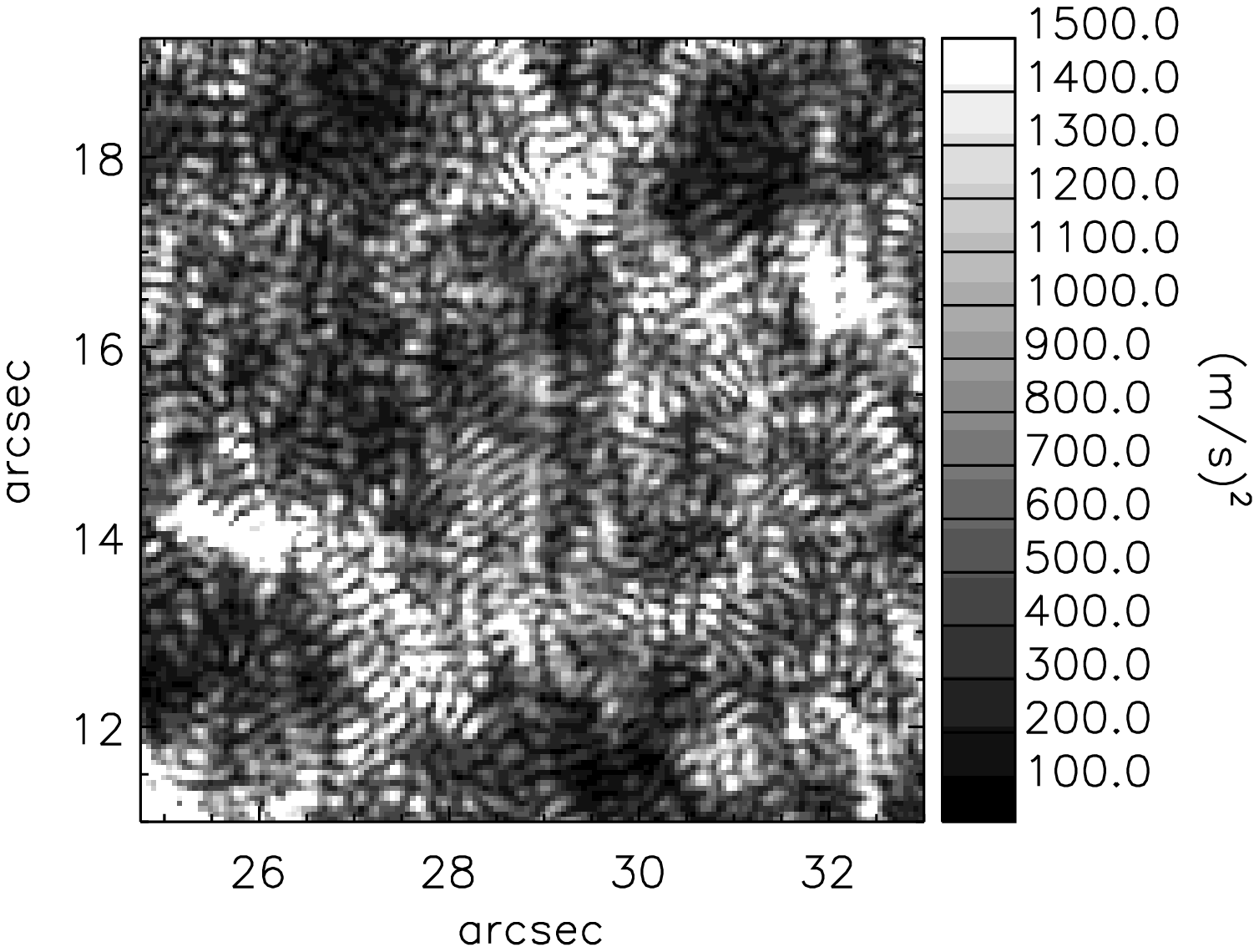}
\caption{Acoustic power maps of the filtered data. White areas are locations on the Sun with high acoustic power. 
{Top}: Average acoustic power of the whole field-of-view. {Bottom}: Average acoustic power of an 8$\arcsec$.25 $\times$ 8$\arcsec$.25 section. The white rectangle {in the top panel}  marks the area which is shown enlarged in the {bottom panel}. {In the bottom panel only values below a threshold of 1500 (m/s)$^2$ are shown to make the acoustic power above the granules visible.}}
\label{fig:acmap}
\end{figure*}

The filtered data allows the excitation and propagation of small-scale acoustic waves to be seen.
For a quick investigation, acoustic power maps are determined by averaging the square of the filtered velocity time series over time and at each pixel.  Figure~\ref{fig:acmap} displays this {temporally averaged} acoustic power of the oscillatory velocity field after filtering.
In this map the granular network is visible, since the intergranular lanes and the borders of the granules exhibit strong acoustic power.
However, there is also significant power present above the granules. This is visible in detail in the enlarged view shown {in the bottom panel} of Fig.~\ref{fig:acmap}. Inspecting the complete series of Doppler maps, the origin of this acoustic power enhancements on top of the granules are oscillations which are either excited in the integranular lanes propagating through the granules or waves which are excited inside splitting or exploding granules.

A time sequence of such an exploding granule is shown in Figs.~\ref{fig:seq} and~\ref{fig:seq2}. 
\begin{figure*}[h]
\epsscale{0.49}
\plotone{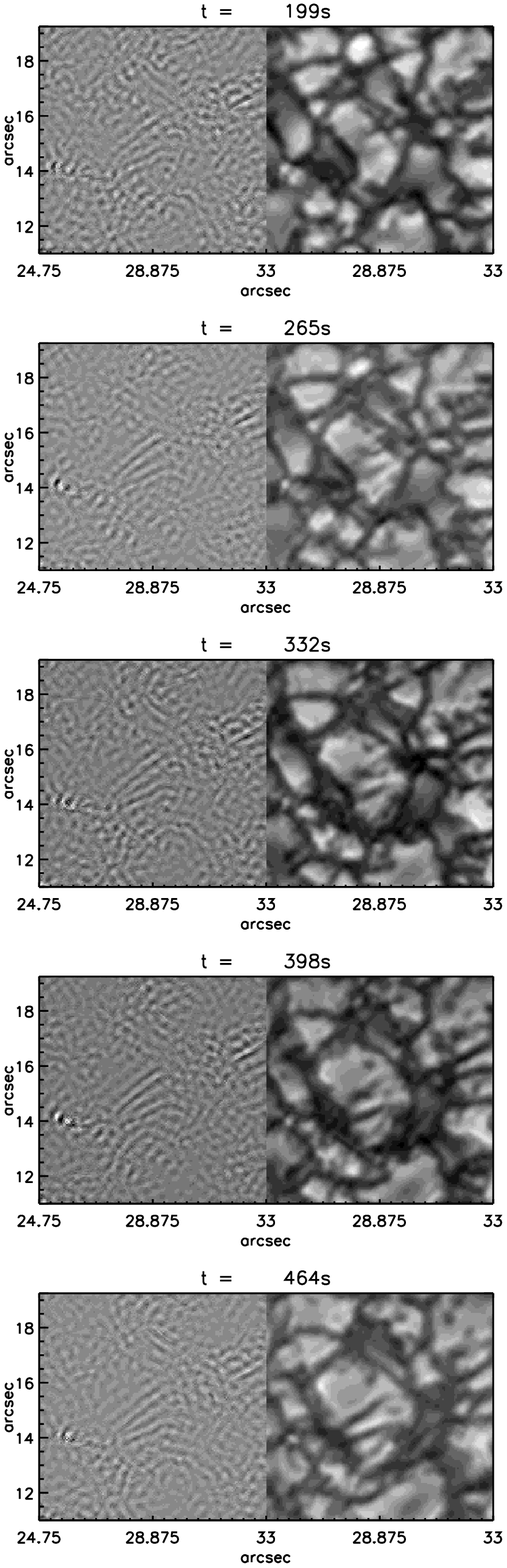}
\caption{Time sequence of the filtered velocity time series (left) and the synchronously recored continuum images of the same area (right). The snapshots are separated by 66\thinspace s. The exploding granule is located in the center. The spatial extent of this area is 8$\arcsec$.3 $\times$ 8$\arcsec$.3. The displayed area corresponds to the map shown in the {bottom} panel of Fig.~\ref{fig:acmap}. The series is continued in Fig.~\ref{fig:seq2}.
}
\label{fig:seq}
\end{figure*}
\begin{figure*}[h]
\epsscale{0.49}
\plotone{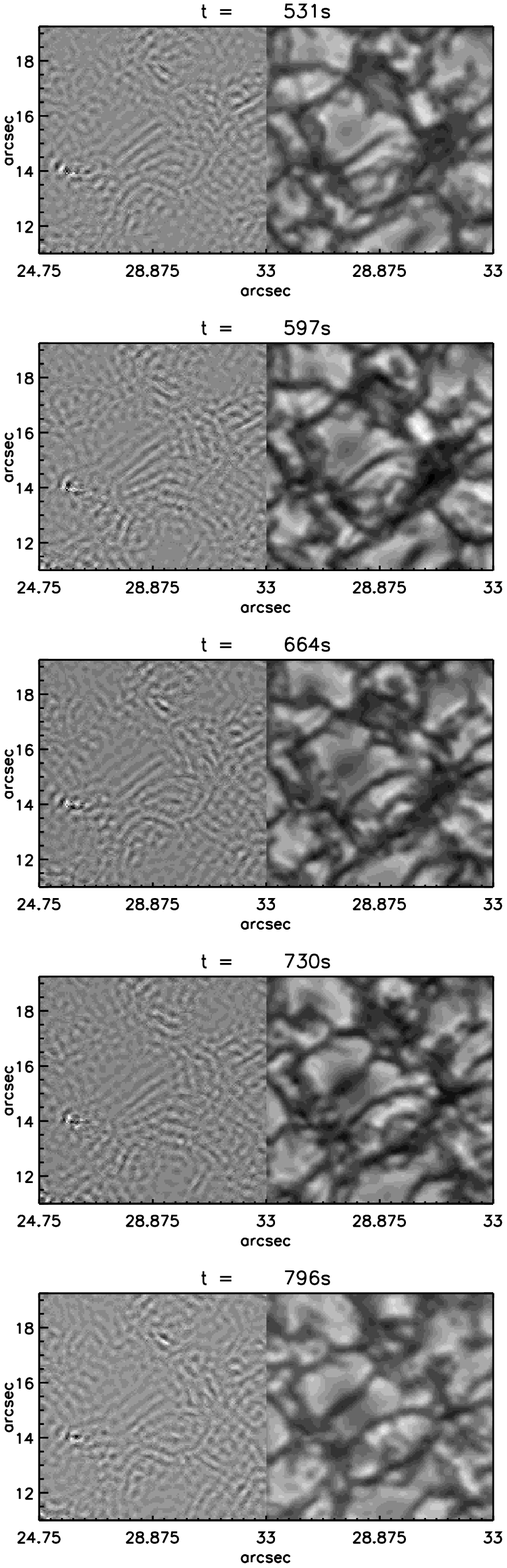}
\caption{Continuation of the time sequence shown in Fig.~\ref{fig:seq}.}
\label{fig:seq2}
\end{figure*}
During the observation the displayed granule continues to grow within the first half of the sequence before it fragments in the second half.
Already at the beginning of the recording (t=199\thinspace s) parallel wave fronts are rippling on top of the granule. Surprisingly, these wave fronts are aligned with an intergranular lane that is forming during the sequence but finally splits the granule 265 \thinspace s later (t=464\thinspace s). Further splittings of the granule occur at later times which seem to excite additional wave fronts on a smaller scale. 
Such events happen in multitude in the recorded time series and not only at this particular location.
Based on a wavelet analysis \citet{bello10} study the distribution of acoustic power in space and time. They find also increased power above small fast-evolving granules, splitting granules, dark dots and forming lanes.

Another finding, which can be observed in all granules of Fig.~\ref{fig:seq} and Fig.~\ref{fig:seq2}, is that the wave amplitude itself is attenuated in the granules. The centers of stable granules show only weak small-scale oscillatory motions. After approximately four wavelengths the high-degree oscillations are damped away. Therefore, waves excited in the intergranular lanes or in the splitting process of granules seem to lose energy while probing a granule. This might explain the comparatively high acoustic power along the borders of the granules.

\section{Conclusions}
\label{sec:conclusions}
We have investigated the properties of small-scale solar $p$-modes as recorded by the IMaX instrument during the flight of the balloon-borne {\sc Sunrise} observatory in June 2009. We analyzed the Doppler velocity and continuum intensity observations to study the potential of such data sets for investigations of the excitation and propagation of high-degree acoustic waves on the Sun. 
Based on two sequences of data, we find that there is a strong relation between increased acoustic $p$-mode power with high {wave number} above granules and the onset of the splitting process that fragments granules. 
The splitting of granules triggers the excitation of waves {several} minutes before the actual splitting becomes visible in continuum radiation. A possible explanation for the excitation of such waves could be related to the formation of new intergranular lanes and connected turbulent motions which {set off} acoustic waves.

Therefore, we conclude that not only the vigorous motions in the established integranular downdrafts are sources for acoustic waves, but also the processes related to the formation of new intergranular lanes.
In future studies, {e.g. based on either data from a second Sunrise flight or from other instruments in space or on ground,} we plan to investigate the excitation process of solar $p$-modes on the basis  of spatially highly resolved data in greater detail.
For a more quantitative analysis we aim to obtain new contiguous observations that cover a longer time period, which shall allow to resolve the ridges of the $p$-modes in the diagnostic diagram.
Such longer time series are needed to measure e.g. the Green's function which is the respose to an ``acoustic event'' on the Sun. This would help determining the statistical properties of the solar oscillation's source function, which is needed as an input to sensitivity kernels as they are calculated, e.g., for local helioseismology.
Furthermore, it remains an open task to study the damping, transmission, and refraction of solar $p$-modes in granules. Observations of such quality could be useful for such investigations and might allow obtaining an acoustic image of the formation and the structure of granules.


\acknowledgements
The German contribution to $Sunrise$ is funded by the Bundesministerium
f\"{u}r Wirtschaft und Technologie through Deutsches Zentrum f\"{u}r Luft-
und Raumfahrt e.V. (DLR), Grant No. 50~OU~0401, and by the Innovationsfond of
the President of the Max Planck Society (MPG). The Spanish contribution has
been funded by the Spanish MICINN under projects ESP2006-13030-C06 and
AYA2009-14105-C06 (including European FEDER funds). The HAO contribution was
partly funded through NASA grant number NNX08AH38G.
M. Roth acknowledges support from the European Helio- and Asteroseismology Network (HELAS) which was funded as Coordination Action under the European Commisssion's Framework Programme 6. This work has been partially supported by the WCU grant No. R31-10016 funded by the Korean Ministry of Education, Science and Technology.



\end{document}